\documentclass[12pt, centerh1]{article}
\usepackage{natbib}
\usepackage[margin=1in]{geometry}
\usepackage{amsmath,natbib,multirow}
\usepackage{amsfonts,mathrsfs,moreverb,lipsum}
\usepackage{graphicx,colonequals}
\usepackage{color}
\usepackage{caption}
\usepackage{mathtools}
\usepackage{pbox}
\usepackage{multirow}
\usepackage{subfigure}
\usepackage{cleveref}
\usepackage{authblk}

\Crefname{figure}{Figure}{Figures}

\newcommand{\btheta}{{\boldsymbol{\theta}}}
\newcommand{\bvtheta}{{\boldsymbol{\vartheta}}}

\newcommand{\vecmu}{\mbox{\boldmath$\mu$}}

\newcommand{\vecx}{\mathbf{x}}
\newcommand{\vecy}{\mathbf{y}}

\newcommand{\vecX}{\mathbf{X}}
\newcommand{\vecU}{\mathbf{U}}

\newcommand{\matsig}{\mathbf\Sigma}
\newcommand{\matPsi}{\mathbf\Psi}

\newcommand{\vecpi}{\mbox{\boldmath$\pi$}}
\newcommand{\vectheta}{\mbox{\boldmath$\theta$}}
\newcommand{\vecphi}{\mbox{\boldmath$\phi$}}

\newcommand{\vecR}{\mathbf{R}}

\newcommand{\vecB}{\mathbf{B}}

\newcommand{\vecY}{\mathbf{Y}}

\newcommand{\vecP}{\mathbf{P}}

\newcommand{\vecalp}{\boldsymbol{\alpha}}
\newcommand{\vecbeta}{\boldsymbol{\beta}}

\allowdisplaybreaks
\title{Multivariate Cluster Weighted Models Using Skewed Distributions}
\author[1]{Michael P.B. Gallaugher}
\author[2]{Salvatore D. Tomarchio$^*$\thanks{$^*$Corresponding author. Email: daniele.tomarchio@unict.it}}
\author[3]{Paul D. McNicholas}
\author[2]{Antonio Punzo}
\affil[1]{Department of Statistical Science, Baylor University, Waco, Texas, USA}
\affil[2]{Department of Economics and Business, University of Catania, Catania, Italy}
\affil[3]{Department of Mathematics and Statistics, McMaster University, Ontario, Canada}

\date{}
\begin{document}
\maketitle{}

\begin{abstract}
Much work has been done in the area of the cluster weighted model (CWM), which extends the finite mixture of regression model to include modelling of the covariates. Although many types of distributions have been considered for both the response and covariates, to our knowledge skewed distributions have not yet been considered in this paradigm. Herein, a family of 24 novel CWMs are considered which allows both the covariates and response variables to be modelled using one of four skewed distributions, or the normal distribution. Parameter estimation is performed using the expectation-maximization algorithm and both simulated and real data are used for illustration.

\end{abstract}

\section{Introduction}
\label{sec:intro}

Clustering is the process of finding underlying group structure in heterogeneous data.
Although many methods exist for clustering, one of the most prevalent in the literature is model-based, and makes use of the $G$ component finite mixture model.
The finite mixture model assumes that the density of a random vector $\vecX$ is
$$
f(\vecx~|~\bvtheta)=\sum_{g=1}^G\pi_gf(\vecx~|~\btheta_g)
$$
where $\pi_g>0$ are the mixing proportions, with $\sum_{g=1}^G\pi_g=1$, $f(\cdot)$ are the component densities parameterized by $\btheta_g$ and $\bvtheta$ contains all the parameters of the model.
As discussed by \cite{mcnicholas16a}, the relationship between the finite mixture model and clustering was initially proposed by \citet{tiedeman55}.
Some years after, \citet{wolfe65} first utilized a Gaussian mixture model for model-based clustering.
Since then, there have been a myriad of contributions to this branch of the literature, mainly considering mixtures of non-Gaussian distributions \citep[a recent review is given by][]{mcnicholas16b}.
Some of these include mixtures of $t$ distributions (\citealp{peel00,andrews11a,andrews12,steane12,lin14}) and power exponential distributions \citep{dang15}, both of which parameterize tail weight and may be useful for modelling data with outliers.
Additionally, many distributions that parameterize both skewness and tail weight have also been proposed. 
These include, but are not limited to, work where mixture components follow a skew-$t$ distribution \citep{lin10,vrbik12,vrbik14,lee14,murray14a,murray14b}, a normal inverse Gaussian distribution \citep{karlis09}, a variance-gamma \citep{mcnicholas17}, a generalized hyperbolic \citep{browne15}, a hidden truncation hyperbolic distribution \citep{murray17b,murray20}, or a skewed power exponential distribution \citep{dang19}.
All of these allow for the modelling of skewed data, which when modelled by a Gaussian distribution has a tendency to over fit the true number of components.

One drawback of the non-Gaussian mixture models mentioned thus far is that they do not typically account for dependencies via covariates.
When there is a clear regression relationship between the variables, important insight can be gained by accounting for functional dependencies between them. 
In such scenarios, the finite mixture of regressions \citep[FMR:][]{desarbo88} may be employed.
As in traditional regression analysis, the FMR model assumes that the covariates are fixed, and therefore the distribution of the covariates is not taken into consideration when performing the cluster analysis. 
Indeed, such a model is also known as finite mixture of regression with fixed covariates.

Unlike the FMR, the cluster weighted model (CWM) offers far more flexibility in that the distribution of the covariates is taken into account.
First introduced by \cite{gershenfeld97}, it is also sometimes referred to as a finite mixture of regression with random covariates.
As discussed in Section~\ref{sec:back}, several CWMs have been introduced in the literature.
Most of them consider a univariate response variable and a set of covariates, modelled by a univariate and a multivariate distribution, respectively.
To our knowledge, only \citet{dang17} consider multiple response variables and covariates, both modelled via multivariate Gaussian distributions.
Herein, we extend this branch of the literature by considering multivariate skewed distributions for both the responses and the covariates.
Specifically the skew-$t$, the generalized hyperbolic, the variance gamma, and the normal inverse Gaussian distributions will be used.
By also considering the Gaussian distribution, we will compose a family of 24 new CWMs, that are flexible enough to consider scenarios in which both the responses and the covariates are skewed, or in which one of the two sets of variables is normally distributed and the other is skewed.

The remainder of this paper is laid out as follows. 
In Section~\ref{sec:back}, a detailed background is given for the cluster weighted model, and the four skewed distributions that will be utilized herein.
Section~\ref{sec:meth} discusses the use of the four skewed distributions in the CWM setting including parameter estimation.
Section~\ref{sec:sim} considers two simulated analyses, in which the parameter recovery and the classification performances for our models are evaluated.
A comparison between FMRs and CWMs is also discussed.
Section~\ref{sec:real} applies our CWMs, along with the Gaussian CWM and the FMRs, to two real datasets.
Lastly, we provide a summary and discuss possible avenues for future work in Section~\ref{sec:end}. 

\section{Background}
\label{sec:back}

\subsection{Cluster Weighted Models}
\label{sec:cwms}

Assume we observe a continuous random response variables $Y_i$ and continuous random covariate vectors $\vecX_i$ of dimension $d$, for a sample of $N$ observations, with $i\in\{1,\ldots,N\}$. 
Also assume the sample can be partitioned into $G$ groups.
In the CWM framework, the joint density of $Y_i$ and $\vecX_i$ can be written as 
\begin{equation}
p(\vecx_i,y_i~|~\bvtheta)=\sum_{g=1}^G\pi_gp_\vecX(\vecx_i~|~\vecphi_g)p_Y(y_i~|~\vecx_i,\vectheta_g),
\label{eq:cwmg}
\end{equation}
where $p_\vecX(\cdot)$ is density function for $\vecX_i$ parameterized by $\vecphi_g$ and $p_Y(\cdot)$ is the density function of $Y_i~|~\vecx_i$ parameterized by $\vectheta_g$.
Note that $\bvtheta=\{\pi_1,\ldots,\pi_G,\vecphi_1,\ldots,\vecphi_G,\btheta_1,\ldots,\btheta_G\}$ represents the set of all parameters.
The CWM is very flexible, and has been thoroughly studied in the literature. 
In its simplest form, it is assumed that $p_\vecX(\cdot)=\Phi_d(\vecx~|~\vecmu_g,\matPsi_g)$ and $p_Y(\cdot)=\Phi_1(y~|~\vecbeta_g'\vecx_i^*,\sigma_g)$, where $\Phi_r(\cdot)$ represents the $r$-dimensional Gaussian density, $\vecbeta$ is a $(d+1)$ dimensional vector of coefficients, and $\vecx_i^*=(1,\vecx_i')'$.

Many extensions of this model have been proposed.
For example, \cite{ingrassia12} propose the use of $t$ distributions for the response and covariates for data with potential outliers. 
Other extensions include  high dimensional covariates \citep{subedi13}, non-linear functional relationships \citep{punzo14c}, detecting outliers using the contaminated normal distribution \citep{punzo14b}, and a general approach that allows various types of response variables as well as covariates of mixed-type \citep{ingrassia15}.
\cite{pocuca20} consider a further extension of \cite{ingrassia15} by further splitting the continuous covariates into Gaussian and non-Gaussian covariates.

Unlike the CWMs just described, \cite{dang17} consider a multivariate response model.
In this case it is assumed that the response $\vecY_i$ is of dimension $p$, so that $p_\vecY(\cdot)=\Phi_p(\vecy~|~\vecB_g'\vecx_i^*,\matsig_g)$, where $\vecB$ is a $(1+d)\times p$ matrix of coefficients.
This multivariate response CWM will be the basis for our family of 24 models, where both or just one of $\vecY_i$ and $\vecX_i$ are allowed to follow a skewed distribution.

\subsection{Generalized Inverse Gaussian Distribution}
\label{subsec:gig}

Before introducing the four skewed distributions which will be used in this paper, the generalized inverse Gaussian distribution is first introduced.
A random variable $Y$ has a generalized inverse Gaussian (GIG) distribution parameterized by $a, b$ and $\lambda$, denoted herein by $\text{GIG}(a,b,\lambda)$, if its probability density function can be written as
$$
f(y|a, b, \lambda)=\frac{\left({a}/{b}\right)^{\frac{\lambda}{2}}y^{\lambda-1}}{2K_{\lambda}(\sqrt{ab})}\exp\left\{-\frac{ay+{b}/{y}}{2}\right\},
$$
where
$$
K_{\lambda}(u)=\frac{1}{2}\int_{0}^{\infty}y^{\lambda-1}\exp\left\{-\frac{u}{2}\left(y+\frac{1}{y}\right)\right\}dy
$$
is the modified Bessel function of the third kind with index $\lambda$. 
Expectations of some functions of a GIG random variable are mathematically tractable,  e.g.:
\begin{equation}
\mathbb{E}(Y)=\sqrt{\frac{b}{a}}\frac{K_{\lambda+1}(\sqrt{ab})}{K_{\lambda}(\sqrt{ab})},
\label{eq:ai}
\end{equation}
\begin{equation}
\mathbb{E}\left({1}/{Y}\right)=\sqrt{\frac{a}{b}}\frac{K_{\lambda+1}(\sqrt{ab})}{K_{\lambda}(\sqrt{ab})}-\frac{2\lambda}{b},
\label{eq:bi}
\end{equation}
\begin{equation}
\mathbb{E}(\log Y)=\log\left(\sqrt{\frac{b}{a}}\right)+\frac{1}{K_{\lambda}(\sqrt{ab})}\frac{\partial}{\partial \lambda}K_{\lambda}(\sqrt{ab}).
\label{eq:ci}
\end{equation}
An alternative parameterization of the generalized inverse Gaussian distribution, as used by \cite{browne15} is given by
\begin{equation}
g(y|\omega,\eta,\lambda)= \frac{\left({w}/{\eta}\right)^{\lambda-1}}{2\eta K_{\lambda}(\omega)}\exp\left\{-\frac{\omega}{2}\left(\frac{w}{\eta}+\frac{\eta}{w}\right)\right\},
\label{eq:I}
\end{equation}
where $\omega=\sqrt{ab}$ and $\eta=\sqrt{a/b}$. 
For notational clarity, we will denote the parameterization given in \eqref{eq:I} by $\text{I}(\omega,\eta,\lambda)$.

\subsection{Skewed Distributions}
\label{subsec:skewed}

Many skewed distributions may be derived by using a normal variance-mean mixture model. 
This model assumes that a random vector $\vecX$ can be written
\begin{equation}
\vecX=\vecmu+W\vecalp+\sqrt{W}\vecU,
\label{eq:vmm}
\end{equation}
where $\vecmu$ is a location parameter, $\vecalp$ is a skewness parameter, $W$ is a positive random variable, and $\vecU\sim\text{N}({\bf 0},\matsig)$, with $\text{N}(\cdot)$ identifying the multivariate normal distribution. 
Herein, we will focus on four different skewed distributions that are special cases of the variance mean mixture model and have been successfully used in model based clustering. 
Namely we will focus on the skew-$t$ (ST), the generalized hyperbolic (GH), the variance gamma (VG) and the normal inverse Gaussian distribution (NIG). 

The $p$-dimensional skew-$t$ distribution, denoted by $\text{ST}(\vecmu,\vecalp,\matsig,\nu)$, arises with $W\sim\text{IGamma}(\nu/2,\nu/2)$, where $\text{IGamma}(\cdot)$ is the inverse gamma distribution.
The resulting density is 
\begin{align*}
f_{\text{ST}}(\vecx~|~\bvtheta)=&\frac{2\left(\frac{\nu}{2}\right)^{\frac{\nu}{2}}\exp\left\{(\vecx-\vecmu)'\matsig^{-1}\vecalp) \right\} }{(2\pi)^{\frac{p}{2}}| \matsig |^{\frac{1}{2}}\Gamma(\frac{\nu}{2})}  \left(\frac{\delta(\vecx;\vecmu,\matsig)+\nu}{\rho(\vecalp,\matsig)}\right)^{-\frac{\nu+p}{4}} \\ & \qquad\qquad\qquad\qquad\times
 K_{-\frac{\nu+p}{2}}\left(\sqrt{\left[\rho(\vecalp,\matsig)\right]\left[\delta(\vecx;\vecmu,\matsig)+\nu\right]}\right),
\end{align*}
where 
$$
\delta(\vecx;\vecmu,\matsig)=(\vecx-\vecmu)'\matsig^{-1}(\vecx-\vecmu), \quad \rho(\vecalp;\matsig)=\vecalp'\matsig^{-1}\vecalp,
$$
and $\nu>0$.

The $p$-dimensional generalized hyperbolic distribution, denoted by $\text{GH}(\vecmu,\vecalp,\matsig,\lambda,\omega)$, arises with $W\sim\text{I}(\omega,1,\lambda)$, and the resulting density is
\begin{align*}
f_{\text{GH}}(\vecx|\bvtheta)=&\frac{\exp\left\{(\vecx-\vecmu)'\matsig^{-1}\vecalp) \right\}}{(2\pi)^{\frac{p}{2}}| \matsig |^{\frac{1}{2}}K_{\lambda}(\omega)}  \left(\frac{\delta(\vecx;\vecmu,\matsig)+\omega}{\rho(\vecalp,\matsig)+\omega}\right)^{\frac{\left(\lambda-\frac{p}{2}\right)}{2}} \\ & \times
 K_{\left(\lambda-{p}/{2}\right)}\left(\sqrt{\left[\rho(\vecalp,\matsig)+\omega\right]\left[\delta(\vecx;\vecmu,\matsig)+\omega\right]}\right),
\end{align*}
$\lambda\in \mathbb{R}$, $\omega\in\mathbb{R}^+$.

The $p$-dimensional variance gamma distribution, denoted by $\text{VG}(\vecmu,\vecalp,\matsig,\gamma)$, arises with $W\sim\text{Gamma}(\gamma,\gamma)$, and the probability density function is
\begin{align*}
f_{\text{VG}}(\vecx|\bvtheta)=&\frac{2\gamma^{\gamma}\exp\left\{(\vecX-\vecmu)\matsig^{-1}\vecalp' \right\}}{(2\pi)^{\frac{p}{2}}| \matsig |^{\frac{1}{2}}\Gamma(\gamma)}  \left(\frac{\delta(\vecx;\vecmu,\matsig)}{\rho(\vecalp,\matsig)+2\gamma}\right)^{\frac{\left(\gamma-{p}/{2}\right)}{2}} \\
&\times  K_{\left(\gamma-\frac{p}{2}\right)}\left(\sqrt{\left[\rho(\vecalp,\matsig)+2\gamma\right]\left[\delta(\vecx;\vecmu,\matsig)\right]}\right),
\end{align*}
where $\gamma\in\mathbb{R}^+$.

Finally, the normal inverse Gaussian, denoted herein by $\text{NIG}(\vecmu,\vecalp,\matsig,\kappa)$, is derived with $W\sim\text{IG}(1,\kappa)$ \citep{ohagan16} where $\text{IG}(\cdot)$ denotes the inverse Gaussian distribution.

\begin{align*}
f_{\text{NIG}}(\vecx|\bvtheta)&=\frac{2\exp\left\{(\vecx-\vecmu)\matsig^{-1}\vecalp)+\kappa\right\}
}{(2\pi)^{\frac{p+1}{2}}| \matsig |^{\frac{1}{2}}}\left(\frac{\delta(\vecx;\vecmu,\matsig)+1}{\rho(\vecalp,\matsig)+\kappa^2}\right)^{-{\left(1+p\right)}/{4}}\\
&\times K_{-{(1+p)}/{2}}\left(\sqrt{\left[\rho(\vecalp,\matsig)+\kappa^2\right]\left[\delta(\vecx;\vecmu,\matsig)+1\right]}\right),
\end{align*}
where $\kappa\in\mathbb{R}^+$.

\section{Methodology}
\label{sec:meth}

\subsection{Cluster Weighted Models With Skewed Distributions}
\label{subsec:cwm_skew}

The cluster weighted model using skewed distributions is now presented. 
For the purposes of this paper, the densities $p_\vecX$ and $p_\vecY$ can be any of the four multivariate skewed distributions previously presented or the multivariate normal distribution. 
In addition, $p_\vecX$ and $p_\vecY$ need not be the same, thus creating a family of 24 new CWMs, plus the completely unconstrained normal CWM of \cite{dang17}.
For notational clarity, each model will be labeled by separating with a ``-'' the acronyms used for $p_\vecX$ and $p_\vecY$, respectively.
For example, the unconstrained normal CWM of \cite{dang17} is herein called N-N CWM.

Recalling the variance mean mixture model in \eqref{eq:vmm}, for an observation in group $g$ the response vector $\vecY_i$, conditional on the covariate vector $\vecX_i$, can be written
$$
\vecY_i|\vecx_i=\vecB_g'\vecx_i^*+V_{ig}{\vecalp_Y}_g+\sqrt{V_{ig}}{\vecU_Y}_{i},
$$
where $\vecB_g$ is a $(1+d)\times p$ matrix of coefficients, $\vecx_i^*=(1,\vecx_i')'$, ${\vecU_Y}_{i}\sim\text{N}({\bf 0},{\matsig_Y}_g)$. 
In the case that $\vecY_i$ is modelled using a multivariate normal distribution then it is assumed $\vecY_i~|~\vecx_i\sim \text{N}(\vecB_g'\vecx_i^*,{\matsig_Y}_g)$. 

If modelling the covariate vector using a skewed distribution, then the random covariate vector, $\vecX_i$, can be written 
$$
\vecX_i=\vecmu_g+W_{ig}{\vecalp_X}_g+\sqrt{W_{ig}}{\vecU_X}_{i},
$$
with ${\vecU_X}_{i}\sim \text{N}({\bf 0},{\matsig_X}_{g})$.
Otherwise, if $\vecX_i$ is modelled using a normal distribution, then $\vecX_i\sim\text{N}(\vecmu_g,{\matsig_X}_g)$.

\subsection{Parameter Estimation}
\label{sec:EM}

The expectation-maximization (EM) algorithm is now utilized for parameter estimation.
For the purposes of this section, we introduce the latent variables $z_{ig}$, where $z_{ig}=1$ if observation $i$ is in group $g$, and $0$ otherwise.
We also introduce the latent variables $W_{ig}$ and $V_{ig}$ if the distributions of $\vecX_i$ and $\vecY_i$, respectively, are skewed. 
The complete data log-likelihood is then 
$$
l(\bvtheta)=l_1(\vecpi)+l_2(\vecphi)+l_3(\btheta),
$$
where $\vecpi=(\pi_1,\ldots,\pi_G)$, $\vecphi=\{\vecphi_1,\ldots,\vecphi_G\}$, $\vectheta=\{\vectheta_1,\ldots,\vectheta_G\}$, and
$$
l_1(\vecpi)=\sum_{g=1}^G\sum_{i=1}^Nz_{ig}\pi_g.
$$
If $\vecX_i$ follows one of the skewed distributions,
\begin{align*}
l_2(\vecphi) & =\sum_{g=1}^G\sum_{i=1}^N\log(h_W(w_{ig}~|~{\boldsymbol \phi_W}_g))+C_X-\frac{1}{2}\sum_{g=1}^G\sum_{i=1}^N z_{ig}[\log(|{\matsig_X}_g|)\\& +(1/w_{ig})(\vecx_i-{\vecmu}_g)'{\matsig_X}_g^{-1}(\vecx_i-{\vecmu}_g)-(\vecx_i-{\vecmu}_g)'{\matsig_X}_g^{-1}{\vecalp_X}_g-{\vecalp_X}_g'{\matsig_X}_g^{-1}(\vecx_i-{\vecmu}_g)\\& +w_{ig}{\vecalp_X}_g'{\matsig_X}_g^{-1}{\vecalp_X}_g],
\end{align*}
where $h_W(\cdot)$ is the density function of $W_{ig}$ parameterized by ${\vecphi_W}_g$, and $C_X$ is constant with respect to the parameters.
On the other hand, if $\vecX_i$ is normally distributed then 
$$
l_2(\vecphi)=C_{XN}-\frac{1}{2}\sum_{g=1}^G\sum_{i=1}^Nz_{ig}[\log(|{\matsig_X}_g|)+(\vecx_i-\vecmu_g)'{\matsig_X}_g^{-1}(\vecx_i-\vecmu_g)],
$$
where $C_{XN}$ is constant with respect to the parameters.

If $\vecY_i$ is distributed according to one of the skewed distributions,
\begin{align*}
l_3(\vectheta) & =\sum_{g=1}^G\sum_{i=1}^N\log(h_V(v_{ig}~|~{\btheta_V}_g))+C_Y-\frac{1}{2}\sum_{g=1}^G\sum_{i=1}^N z_{ig}[\log(|{\matsig_Y}_g|)\\& +(1/v_{ig})(\vecy_i-{\vecB_g'\vecx_i^*})'{\matsig_Y}_g^{-1}(\vecy_i-{\vecB_g'\vecx_i^*})-(\vecy_i-{\vecB_g'\vecx_i^*})'{\matsig_Y}_g^{-1}{\vecalp_Y}_g-{\vecalp_Y}_g'{\matsig_Y}_g^{-1}(\vecy_i-{\vecB_g'\vecx_i^*})\\& +v_{ig}{\vecalp_Y}_g'{\matsig_Y}_g^{-1}{\vecalp_Y}_g],
\end{align*}
where $h_V(\cdot)$ is the density function of $V_{ig}$ parameterized by ${\btheta_V}_g$, and $C_Y$ is a constant with respect to the parameters.
If $\vecY_i$ is normally distributed,
$$
l_3(\vectheta)=C_{YN}-\frac{1}{2}\sum_{g=1}^G\sum_{i=1}^Nz_{ig}[\log(|{\matsig_Y}_g|)+(\vecx_i-\vecmu_g)'{\matsig_Y}_g^{-1}(\vecx_i-\vecmu_g)],
$$
where $C_{YN}$ is constant with respect to the parameters.
After initialization, the EM algorithm proceeds as follows.\\
\noindent {\bf E Step:} Update the group memberships $z_{ig}$ given by 
$$
\hat{z}_{ig}=\frac{\pi_gp(\vecy_i,\vecx_i~|~\hat{\vecphi}_g,\hat{\btheta}_g)}{\sum_{h=1}^G\pi_hp(\vecy_i,\vecx_i~|~\hat{\vecphi}_h,\hat{\btheta}_h)}.
$$
If the distribution of $\vecX_i$ is skewed, then in addition, the following values need to be updated
\begin{equation*}
\begin{split}
a_{ig}&\colonequals\mathbb{E}[W_{ig}~|~z_{ig}=1,\vecx_i,\hat{\vecphi}_g]\\
b_{ig}&\colonequals\mathbb{E}[1/W_{ig}~|~z_{ig}=1,\vecx_i,\hat{\vecphi}_g]\\
c_{ig}&\colonequals\mathbb{E}[\log(W_{ig})~|~z_{ig}=1,\vecx_i,\hat{\vecphi}_g]\\
\end{split}
\end{equation*} 
If the distribution of $\vecY_i$ is skewed, then the following values are also updated
\begin{equation*}
\begin{split}
k_{ig}&\colonequals\mathbb{E}[V_{ig}~|~z_{ig}=1,\vecy_i,\hat{\btheta}_g]\\
m_{ig}&\colonequals\mathbb{E}[1/V_{ig}~|~z_{ig}=1,\vecy_i,\hat{\btheta}_g]\\
n_{ig}&\colonequals\mathbb{E}[\log(V_{ig})~|~z_{ig}=1,\vecy_i,\hat{\btheta}_g]\\
\end{split}
\end{equation*} 
These updates are dependent upon the distribution; however, we have the following properties of the conditional distributions of the latent variables for each of the four distributions.
\begin{align*}
W_{ig}^{\text{ST}}~|~\vecx_i, z_{ig}=1&\sim \text{GIG}\left(\rho({\vecalp_X}_g,{\matsig_X}_g),\delta(\vecx_i;\vecmu_g,{\matsig_X}_g)+{\nu_{X}}_g,-({\nu_{X}}_g+p)/2\right)\\
W_{ig}^{\text{GH}}~|~\vecx_i, z_{ig}=1&\sim \text{GIG}\left(\rho({\vecalp_X}_g,{\matsig_X}_g)+{\omega_{X}}_{g},\delta(\vecx_i;\vecmu_g,{\matsig_X}_g)+{\omega_{X}}_g,{\lambda_{X}}_g-{p}/{2}\right)\\
W_{ig}^{\text{VG}}~|~\vecx_i, z_{ig}=1&\sim \text{GIG}\left(\rho({\vecalp_X}_g,{\matsig_X}_g)+2{\gamma_{X}}_g,\delta(\vecx_i;\vecmu_g,{\matsig_X}_g),{\gamma_{X}}_g-{p}/{2}\right)\\
W_{ig}^{\text{NIG}}~|~\vecx_i, z_{ig}=1&\sim \text{GIG}\left(\rho({\vecalp_X}_g,{\matsig_X}_g)+{\kappa^2_{X}}_g,\delta(\vecx_i;\vecmu_g,{\matsig_X}_g)+1,-{(1+p)}/{2}\right)
\end{align*}
and 
\begin{align*}
V_{ig}^{\text{ST}}~|~\vecx_i, \vecy_i, z_{ig}=1&\sim \text{GIG}\left(\rho({\vecalp_Y}_g,{\matsig_Y}_g),\delta(\vecy_i;\vecB_g'\vecx_i,{\matsig_Y}_g)+{\nu_{Y}}_g,-({\nu_{Y}}_g+p)/2\right)\\
V_{ig}^{\text{GH}}~|~\vecx_i, \vecy_i, z_{ig}=1&\sim \text{GIG}\left(\rho({\vecalp_Y}_g,{\matsig_Y}_g)+{\omega_{Y}}_g,\delta(\vecy_i;\vecB_g'\vecx_i,{\matsig_Y}_g)+{\omega_{Y}}_g,{\lambda_{Y}}_g-{p}/{2}\right)\\
V_{ig}^{\text{VG}}~|~\vecx_i, \vecy_i, z_{ig}=1&\sim \text{GIG}\left(\rho({\vecalp_Y}_g,{\matsig_Y}_g)+2{\gamma_{Y}}_g,\delta(\vecy_i;\vecB_g'\vecx_i,{\matsig_Y}_g),{\gamma_{Y}}_g-{p}/{2}\right)\\
V_{ig}^{\text{NIG}}~|~\vecx_i, \vecy_i, z_{ig}=1&\sim \text{GIG}\left(\rho({\vecalp_Y}_g,{\matsig_Y}_g)+{\kappa^2_{Y}}_g,\delta(\vecy_i;\vecB_g'\vecx_i,{\matsig_Y}_g)+1,-{(1+p)}/{2}\right).
\end{align*}
Therefore, all of the required expectations can be calculated using \eqref{eq:ai}--\eqref{eq:ci}.

\noindent {\bf M Step:} In the M step, we update all of the parameters. 
Specifically, the parameters for the distribution of $\vecX_i$ are updated as follows. 
If $\vecX_i$ is normally distributed then
\begin{equation*}
\hat{\vecmu}_g=\frac{1}{T_g}\sum_{g=1}^G\hat{z}_{ig}\vecx_i, \qquad  \hat{{\matsig}}_{X_g}=\frac{1}{T_g}\sum_{g=1}^G\hat{z}_{ig}(\vecx_i-\hat{\vecmu}_g)(\vecx_i-\hat{\vecmu}_g)',
\end{equation*}
with $T_g=\sum_{i=1}^N\hat{z}_{ig}$.
On the other hand, if $\vecX_i$ follows one of the skewed distributions, then we have the following updates for the related parameters
\begin{equation*}
\hat{\vecmu}_g=\frac{\sum_{i=1}^N\hat{z}_{ig}\vecx_i\left(\overline{a}_gb_{ig}-1\right)}{\sum_{i=1}^N\hat{z}_{ig}\overline{a}_gb_{ig}-T_g},\qquad
\hat{{\vecalp}}_{X_g}=\frac{\sum_{i=1}^N\hat{z}_{ig}\vecx_i\left(\overline{b}_g-b_{ig}\right)}{\sum_{i=1}^N\hat{z}_{ig}\overline{a}_gb_{ig}-T_g},
\end{equation*}
where $\overline{a}_g=(1/T_g)\sum_{i=1}^N\hat{z}_{ig}a_{ig}$ and $\overline{b}_g=(1/T_g)\sum_{i=1}^N\hat{z}_{ig}b_{ig}$.
The update for ${\matsig_X}_g$ in this case is
$$
\hat{{\matsig}}_{X_g}=\frac{1}{T_g}\sum_{i=1}^N\hat{z}_{ig}\left[b_{ig}(\vecx_i-\hat{\vecmu}_g)(\vecx_i-\hat{\vecmu}_g)'-(\vecx_i-\hat{\vecmu}_g)\hat{{\vecalp}}_{X_g}'-\hat{{\vecalp}}_{X_g}(\vecx_i-\hat{\vecmu}_g)'+a_{ig}\hat{{\vecalp}}_{X_g}\hat{{\vecalp}}_{X_g}'\right]
$$

If $\vecY_i$ is modelled using a multivariate normal distribution, the update for $\vecB_g$ is given by 
$$
\hat{\vecB}_g=\left(\sum_{i=1}^N\hat{z}_{ig}\vecx_i^*{\vecx_i^*}'\right)^{-1}\left(\sum_{i=1}^N\hat{z}_{ig}\vecx_i^*{\vecy_i}'\right),
$$
and the update for ${\matsig_Y}_g$ is
$$
\hat{{\matsig}}_{Y_g}=\frac{1}{T_g}\sum_{i=1}^N\hat{z}_{ig}(\vecy_i-\hat{\vecB}_g'\vecx_i)(\vecy_i-\hat{\vecB}_g'\vecx_i)'.
$$
If, however, $\vecY_i$ follows one of the skewed distributions, the updates for $\vecB_g$ and $\vecalp_{Y_g}$ are given by
$$
\hat{\vecB}_g=\vecP_g^{-1}\vecR_g, \qquad \hat{{\vecalp}}_{Y_g}=\frac{1}{T_g\overline{k}_g}\left(\sum_{i=1}^N\hat{z}_{ig}\vecy_i-\vecR_g'\vecP_g^{-1}\sum_{i=1}^N\hat{z}_{ig}\vecx_i^*\right),
$$
where 
$$
\vecP_g=\sum_{i=1}^N\hat{z}_{ig}m_{ig}\vecx_i^*{\vecx_i^*}'-\frac{1}{T_g\overline{k}_g}\left(\sum_{i=1}^N\hat{z}_{ig}\vecx_i^*\right)\left(\sum_{i=1}^N\hat{z}_{ig}{\vecx_i^*}'\right)
$$
and 
$$
\vecR_g=\sum_{i=1}^N\hat{z}_{ig}m_{ig}\vecx_i^*{\vecy_i}'-\frac{1}{T_g\overline{k}_g}\left(\sum_{i=1}^N\hat{z}_{ig}\vecx_i^*\right)\left(\sum_{i=1}^N\hat{z}_{ig}{\vecy_i}'\right),
$$
with $\overline{k}_g=(1/T_g)\sum_{i=1}^N\hat{z}_{ig}k_{ig}$.
The update for ${\matsig_Y}_g$ in this case is
\begin{equation*}\begin{split}
\hat{{\matsig}}_{Y_g} & =\\&\frac{1}{T_g}\sum_{i=1}^N\hat{z}_{ig}\left[m_{ig}\left(\vecy-\hat{\vecB}_g'\vecx_i^*\right)\left(\vecy-\hat{\vecB}_g'\vecx_i^*\right)'-\left(\vecy-\hat{\vecB}_g'\vecx_i^*\right)\hat{{\vecalp}}_{Y_g}'-\hat{{\vecalp}}_{Y_g}\left(\vecy-\hat{\vecB}_g'\vecx_i^*\right)' +k_{ig}\hat{{\vecalp}}_{Y_g}\hat{{\vecalp}}_{Y_g}'\right].
\end{split}\end{equation*}

Finally, if either $\vecX_i$ or $\vecY_i$ follows one of the skewed distributions, then there are the additional concentration and, in the case of the generalized hyperbolic distribution, the index parameters that need to be updated. 
The updates for each distribution are now given.

\subsubsection*{Skew-$t$ Distribution}
\label{sec:st}

In the case of the skew-$t$ distribution, we need to update the degrees of freedom, $\nu_g$. This update cannot be obtained in closed form, and thus needs to be performed numerically. When $\vecX_i$ is considered, the update $\nu_{X_g}^{(t+1)}$ is obtained by solving \eqref{eq:nugup} for $\nu_{X_g}$,
\begin{equation}
\log\left(\frac{\nu_{X_g}}{2}\right)+1-\varphi\left(\frac{\nu_{X_g}}{2}\right)-\frac{1}{T_g}\sum_{i=1}^N\hat{z}_{ig}^{(t+1)}(b^{(t+1)}_{ig}+c^{(t+1)}_{ig})=0,
\label{eq:nugup}
\end{equation}
where $\varphi(\cdot)$ denotes the digamma function.
When $\vecY_i$ is considered, the update for $\nu_{Y_g}^{(t+1)}$ is obtained via~\eqref{eq:nugup}, after the replacement of $\nu_{X_g}$, $b_{ig}$ and $c_{ig}$ with $\nu_{Y_g}$, $m_{ig}$ and $n_{ig}$, respectively.

\subsubsection*{Generalized Hyperbolic Distribution}
\label{sec:gh}

For the generalized hyperbolic distribution, we would update $\lambda_g$ and $\omega_g$.
These updates are derived from \cite{browne15}, and rely on the log convexity of $K_{s}(t)$, \cite{baricz10}, in both $s$ and $t$.
For notational purposes in this section, the superscript $t$ denotes the update at the previous iteration.
The resulting updates, when $\vecX_i$ is considered, are
\begin{align}
\hat{\lambda}_{X_g}^{(t+1)}&=\bar{c}_g\hat{\lambda}_{X_g}^{(t)}\left[\left.\frac{\partial}{\partial s}\log(K_{s}(\hat{\omega}_{X_g}^{(t)}))\right|_{s=\hat{\lambda}_{X_g}^{(t)}}\right]^{-1} \label{eq:lamup}\\
\hat{\omega}_{X_g}^{(t+1)}&=\hat{\omega}_{X_g}^{(t)}-\left[\left.\frac{\partial}{\partial s}q(\hat{\lambda}_{X_g}^{(t+1)},s)\right|_{s=\hat{\omega}_{X_g}^{(t)}}\right]\left[\left.\frac{\partial^2}{\partial s^2}q(\hat{\lambda}_{X_g}^{(t+1)},s)\right|_{s=\hat{\omega}_{X_g}^{(t)}}\right]^{-1}
\label{eq:omup}
\end{align}
where the derivative in \eqref{eq:lamup} is calculated numerically, 
$$
q(\lambda_{X_g},\omega_{X_g})=\sum_{i=1}^Nz_{ig}\left[\log(K_{\lambda_{X_g}}(\omega_{X_g}))-\lambda_{X_g}\overline{c}_{ig}-\frac{1}{2}\omega_{X_g}\left(\overline{a}_{ig}+\overline{b}_{ig}\right)\right],
$$
and $\bar{c}_g=({1}/{T_g})\sum_{i=1}^N\hat{z}_{ig}c_{ig}$.
When $\vecY_i$ is considered, $\omega_{X_g}, \lambda_{X_g}$, $\bar{a}_g, \bar{b}_g$, and $\bar{c}_g$ are replaced with $\omega_{Y_g}, \lambda_{Y_g}$, $\bar{k}_g, \bar{m}_g$, and $\bar{n}_g$, respectively.

\subsubsection*{Variance-Gamma Distribution}
\label{sec:vg}

For the variance-gamma, the update for $\gamma_g$, like the generalized hyperbolic case, cannot be obtained in closed form. 
For the $\vecX_i$, this update is obtained by solving \eqref{eq:gammup} for $\gamma_{X_g}$
\begin{equation}
\log\gamma_{X_g}+1-\varphi(\gamma_{X_g})+\bar{c}_g-\bar{a}_g=0.
\label{eq:gammup}
\end{equation}
When $\vecY_i$ is considered, $\gamma_{X_g}$, $\bar{a}_g$ and $\bar{c}_g$ are replaced with $\gamma_{Y_g}$, $\bar{k}_g$ and $\bar{n}_g$, respectively.

\subsubsection*{Normal Inverse Gaussian Distribution}
\label{sec:nig}
When we consider $\vecX_i$, the update of $\kappa_g$ has the following closed form 
$$
\kappa_{X_g}^{(t+1)}=\frac{1}{\overline{a}^{(t+1)}_g}.
$$
If $\vecY_i$ is considered, we replace $\kappa_{X_g}$ and $\bar{a}_g$ with $\kappa_{Y_g}$ and $\bar{k}_g$, respectively.

\subsection{Initialization of the Algorithm}
\label{sec:init}

To initialize the EM algorithm, we followed the approach discussed in \citet{dang17}.
Specifically, the $z_{ig}$ are initialized in two different ways: 10 times using a random soft initialization and once with a $k$-means hard initialization.
Therefore, for each $G$, the algorithms are run 11 times, and the solution producing the highest log-likelihood value is chosen.
Notice that, for the $k$-means initialization, the initial $z_{ig}$ are selected from the best k-means clustering results from 10 random starting values, and it is implemented by using the \texttt{kmeans()} function of the R statistical software \citep{R19}.
For a better comparability with the competing models, the same $z_{ig}$ are also used to initialize the FMRs considered herein.

\section{Simulated Data Analyses}
\label{sec:sim}
In this section, several aspects related to our models are analyzed.
First, in Section~\ref{sec:pr} the parameter recovery, classification performance, and selection of the number of groups is discussed.
Classification performance is evaluated by computing the adjusted Rand index (ARI; \citealp{hubert85}), which calculates the agreement between the true classification and the one predicted by the model. An ARI of 1 indicates perfect agreement between the two partitions, whereas the expected value of the ARI under random classification is 0.
The Bayesian information criterion (BIC; \citealp{schwarz78}) is used to assess the selection of the true number of groups.

In Section~\ref{sec:cc}, our models and their competitors are tested under different scenarios. 
Specifically, a comparison between the CWMs and the FMRs is conducted. In addition, a comparison between our CWMs and the N-N CWM is performed, and the capability of the BIC in detecting the data generating model is evaluated.

\subsection{Parameter Recovery and Classification Evaluation}
\label{sec:pr}

Because of the high number of CWMs introduced in this manuscript, we will focus our attention on four of the 24 novel CWMs.
Specifically, we analyze four models that can cover the following different scenarios:
\begin{enumerate}
\item $p_{\vecX}$ and $p_{\vecY}$ are the same skewed density;
\item $p_{\vecX}$ and $p_{\vecY}$ are different skewed densities,
\item $p_{\vecX}$ is skewed and $p_{\vecY}$ is normal;
\item $p_{\vecX}$ is normal and $p_{\vecY}$ is skewed.
\end{enumerate}
As illustrative examples, we consider the (1) GH-GH CWM, (2) VG-NIG CWM, (3) ST-N CWM and (4) N-NIG CWM.
Note that the models are chosen so that all the distributions considered in this manuscript are incorporated in some capacity.

According to the CWM literature (see, e.g.~\citealp{punzo14c,ingrassia15,punzo15b,punzo14b}), and because of the high number of parameters that should be otherwise reported, we limit our analysis to the recovery of the regression coefficients.
We consider the case with $p=2$, $r=3$ and $N=400$.
The parameters used to generate the data, and are equal for the four CWMs, are displayed in~\tablename~\ref{tab:param}.
The additional parameters, specific for each model, are: $\omega_{X_1}=4.00$, $\omega_{X_2}=10.00$, $\omega_{Y_1}=10.00$, $\omega_{Y_2}=4.00$ and $\lambda_{X_1}=\lambda_{X_2}=\lambda_{Y_1}=\lambda_{Y_2}=0.30$ for the GH-GH CWM, $\gamma_{X_1}=\gamma_{Y_1}=4.00$, $\gamma_{X_2}=20.00$ and $\gamma_{Y_2}=10.00$ for the VG-NIG CWM, $\nu_{X_1}=\nu_{X_2}=7.00$ for the ST-N CWM, $\kappa_{Y_1}=4.00$ and $\kappa_{Y_2}=10.00$ for the N-NIG CWM.
\begin{table}[!ht]
    \centering
\caption{Common parameters between the four CWMs used to generate the simulated datasets.}
\label{tab:param}       
       \begin{tabular}{lcc}
				\hline\noalign{\smallskip}
	Parameter  & Group 1 & Group 2 \\
	      \hline\noalign{\smallskip}
	$\pi_g$	& 0.50 & 0.50 \\		
	$\vecmu_g$	 & $(0.00,0.00,0.00)'$  & $(3.00,3.00,3.00)'$    \\
	$\vecalp_{X_g}$ & $(2.00,2.00,2.00)'$  & $(-3.00,-3.00,-3.00)'$ \\
	$\matsig_{X_g}$  & $\begin{pmatrix}  1.00 & 0.10 & 0.20  \\
                                  0.10 & 3.00 & 0.10  \\
																	0.20 & 0.10 & 2.00  \end{pmatrix}$ & $\begin{pmatrix}  1.00 & 0.10 & 0.10  \\
                                                                                         0.10 & 1.00 & 0.20  \\
		 															                                                       0.10 & 0.20 & 1.00  \end{pmatrix}$ \\[+10mm]
  $\vecB_g$   & $\begin{pmatrix}  -6.00 & 1.00  \\
                                  -1.50 & -1.50 \\
																	-0.50 & -1.50 \\
																	 2.50 &	 1.50 \end{pmatrix}$ & $\begin{pmatrix}  10.00 & -7.50  \\
                                                                                   -6.00 &  4.00 \\
																																									  4.00 &  5.50 \\
																																									 -3.50 & -3.00 \end{pmatrix}$\\
	$\vecalp_{Y_g}$ & $(2.00,-2.00)'$  & $(-2.00,2.00)'$ \\
	$\matsig_{Y_g}$ & $\begin{pmatrix}  1.00 & 0.20  \\
                                  0.20 & 1.00  \end{pmatrix}$ & $\begin{pmatrix}  1.00 & 0.30  \\
                                                                                  0.30 & 1.00  \end{pmatrix}$ \\
	\noalign{\smallskip}\hline
        \end{tabular}
\end{table}

For each of the four CWMs, 100 datasets are generated and the corresponding model is fitted with $G=2$.
The average and the standard deviation of the regression coefficient estimates of each model, over the 100 datasets, are reported in~\tablename~\ref{tab:res_par}.
\begin{table}[!ht]
    \centering
\caption{Average ($\overline{\vecB}_g$) and standard deviation ($\sigma_{\vecB_g}$) of the regression coefficients estimates over 100 datasets for each CWM.}
\label{tab:res_par}       
       \begin{tabular}{lcc|cc}
				\hline\noalign{\smallskip}
CWM & \multicolumn{2}{c}{$\overline{\vecB}_g$} & \multicolumn{2}{c}{$\sigma_{\vecB_g}$} \\
    & Group 1 & Group 2 & Group 1 & Group 2  \\
	      \hline\noalign{\smallskip}
GH-GH  & $\begin{pmatrix*}[r]  -6.26  & 1.12  \\
																-1.49 & -1.51 \\
																-0.49 & 1.50  \\
																2.49  & 1.51 \end{pmatrix*}$ & $\begin{pmatrix*}[r] 10.73 & -8.14 \\
																																								-6.02 & 4.00  \\
																																								4.01  & 5.50  \\
																																								-3.50 & -3.00 \end{pmatrix*}$ & $\begin{pmatrix*}[r]  0.74 & 0.78 \\
                               0.06 & 0.06 \\
                               0.05 & 0.04 \\
                               0.05 & 0.06\end{pmatrix*}$ & $\begin{pmatrix*}[r]  0.87 & 0.82 \\
                                                                             0.09 & 0.07 \\
                                                                             0.08 & 0.08 \\
                                                                             0.09 & 0.08 \end{pmatrix*}$\\ [+14mm]
VG-NIG  & $\begin{pmatrix*}[r]  -6.30 & 1.32  \\
																-1.50 & -1.50 \\
																-0.50 & 1.50  \\
																2.50  & 1.50 \end{pmatrix*}$ & $\begin{pmatrix*}[r]  9.96  & -7.44 \\
																																								-6.00 & 4.00  \\
																																								4.00  & 5.50  \\
																																								-3.50 & -3.00 \end{pmatrix*}$ & $\begin{pmatrix*}[r]  0.41 & 0.40 \\
                                0.03 & 0.03 \\
                                0.02 & 0.02 \\
                                0.03 & 0.03 \end{pmatrix*}$ & $\begin{pmatrix*}[r]  0.17 & 0.16 \\
																																							 0.02 & 0.02 \\
																																						   0.02 & 0.02 \\
																																						   0.02 & 0.02 \end{pmatrix*}$\\ [+14mm]
ST-N  & $\begin{pmatrix*}[r]  -6.00 & 1.00  \\
															-1.50 & -1.49 \\
															-0.50 & 1.50  \\
															2.50  & 1.49 \end{pmatrix*}$ & $\begin{pmatrix*}[r]  10.00 & -7.50 \\
																																							-6.00 & 4.00  \\
																																							4.00  & 5.50  \\
																																							-3.51 & -3.00 \end{pmatrix*}$ & $\begin{pmatrix*}[r]  0.13 & 0.13 \\
                              0.05 & 0.05 \\
                              0.03 & 0.03 \\
                              0.04 & 0.04\end{pmatrix*}$ & $\begin{pmatrix*}[r]  0.08 & 0.08 \\
                                                                            0.06 & 0.05 \\
                                                                            0.06 & 0.06 \\
                                                                            0.05 & 0.05 \end{pmatrix*}$\\ [+14mm]
N-NIG  & $\begin{pmatrix*}[r]  -5.98 & 1.03  \\
															 -1.50 & -1.50 \\
															 -0.50 & 1.50  \\
															 2.50  & 1.50 \end{pmatrix*}$ & $\begin{pmatrix*}[r]  9.93  & -7.43 \\
																																							 -6.00 & 4.00  \\
																																							 3.99  & 5.50  \\
																																							 -3.50 & -3.00 \end{pmatrix*}$ & $\begin{pmatrix*}[r]  0.17 & 0.17 \\
                               0.03 & 0.04 \\
                               0.02 & 0.02 \\
                               0.03 & 0.03\end{pmatrix*}$ & $\begin{pmatrix*}[r]  0.18 & 0.18 \\
                                                                             0.02 & 0.02 \\
                                                                             0.02 & 0.02 \\
                                                                             0.02 & 0.02 \end{pmatrix*}$\\ [+14mm]
\noalign{\smallskip}\hline
        \end{tabular}
\end{table}

Overall, the average estimates for the coefficient matrices are very close to their true values; however, the estimates for the intercepts are a little less accurate.
From the analysis of the standard deviations, we can see those related to the GH-GH CWM are slightly higher than the other CWMs, which may be due to the added complexity of the index parameter.

The average ARI estimates, and the classification performance, is very good for all models considered (\tablename~\ref{tab:classbic}). Finally, we note that when fitting the models for $G=1,2,3$ on the same 100 datasets, the correct number of groups ($G=2$) is always selected by the BIC.
\begin{table}[!ht]
    \centering
\caption{Average ARI values ($\overline{\text{ARI}}$) computed over the 100 datasets for the CWMs.}
\label{tab:classbic}       
       \begin{tabular}{lccc|c}
				\hline\noalign{\smallskip}
	CWM  & $\overline{\text{ARI}}$ \\
	      \hline\noalign{\smallskip}
	GH-GH  & 0.96  \\		
	VG-NIG & 1.00  \\	
  ST-N   & 0.99  \\
	N-NIG  & 1.00  \\	
	\noalign{\smallskip}\hline
        \end{tabular}
\end{table}

\subsection{Comparison Between CWMs and FMRs}
\label{sec:cc}

\begin{figure}[!ht]
\centering
\subfigure[\label{fig:CWM1}]
{\includegraphics[width=0.495\textwidth]{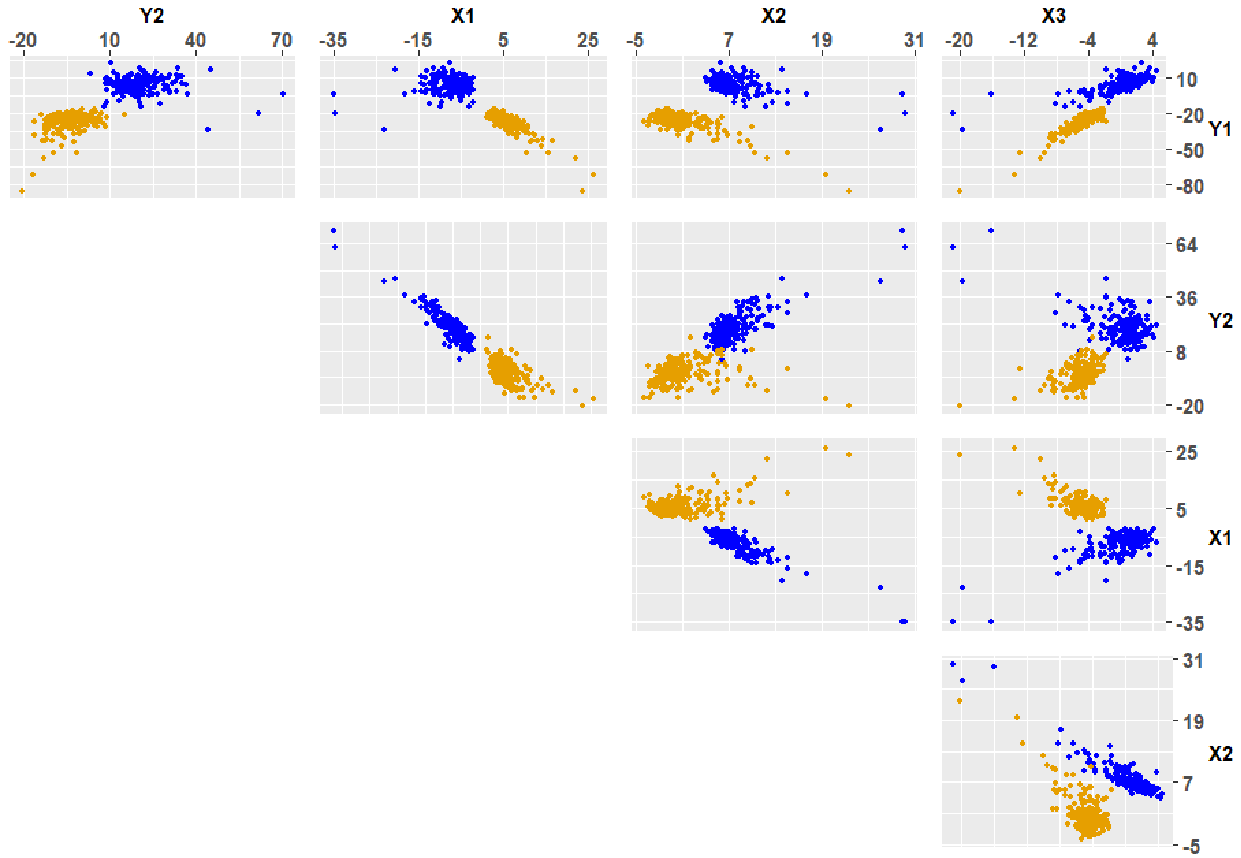}}
\subfigure[\label{fig:CWM2}]
{\includegraphics[width=0.495\textwidth]{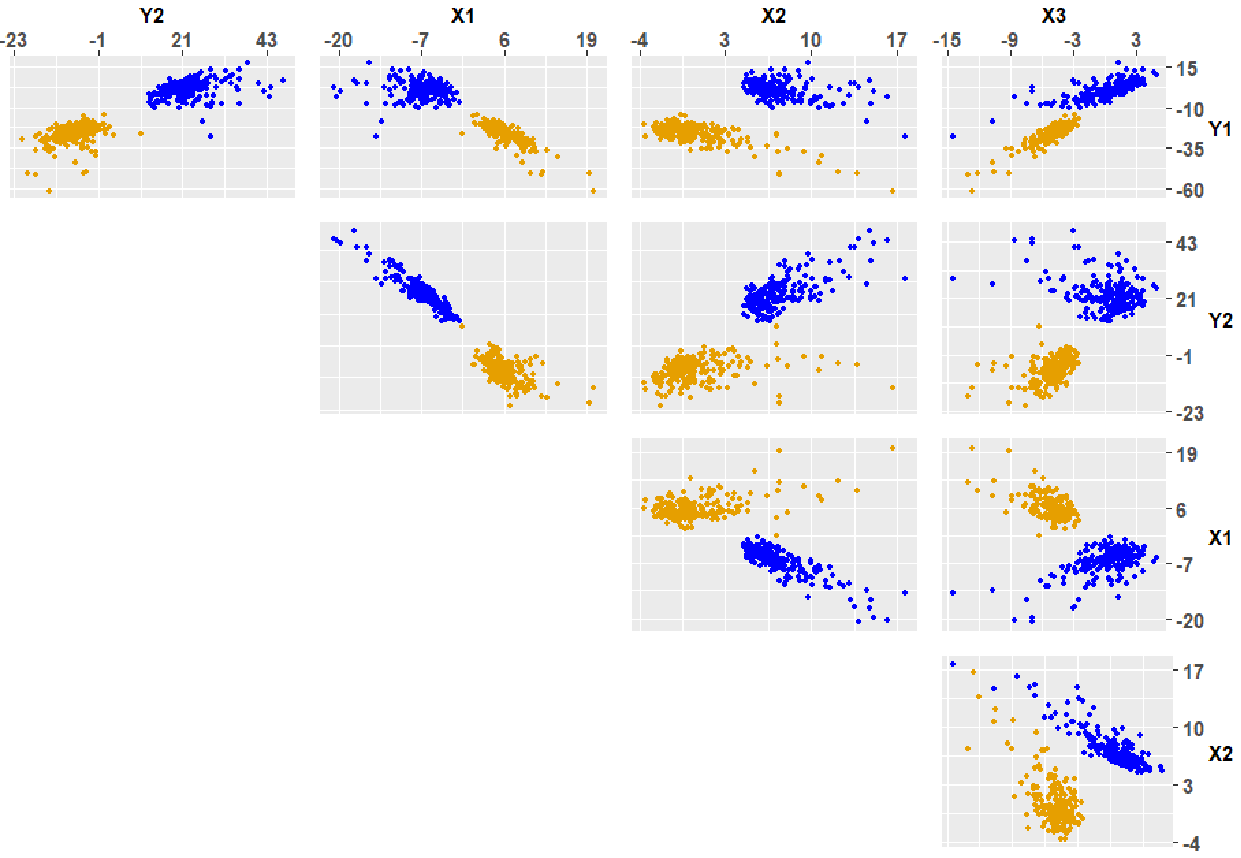}}
\subfigure[\label{fig:CWM3}]
{\includegraphics[width=0.495\textwidth]{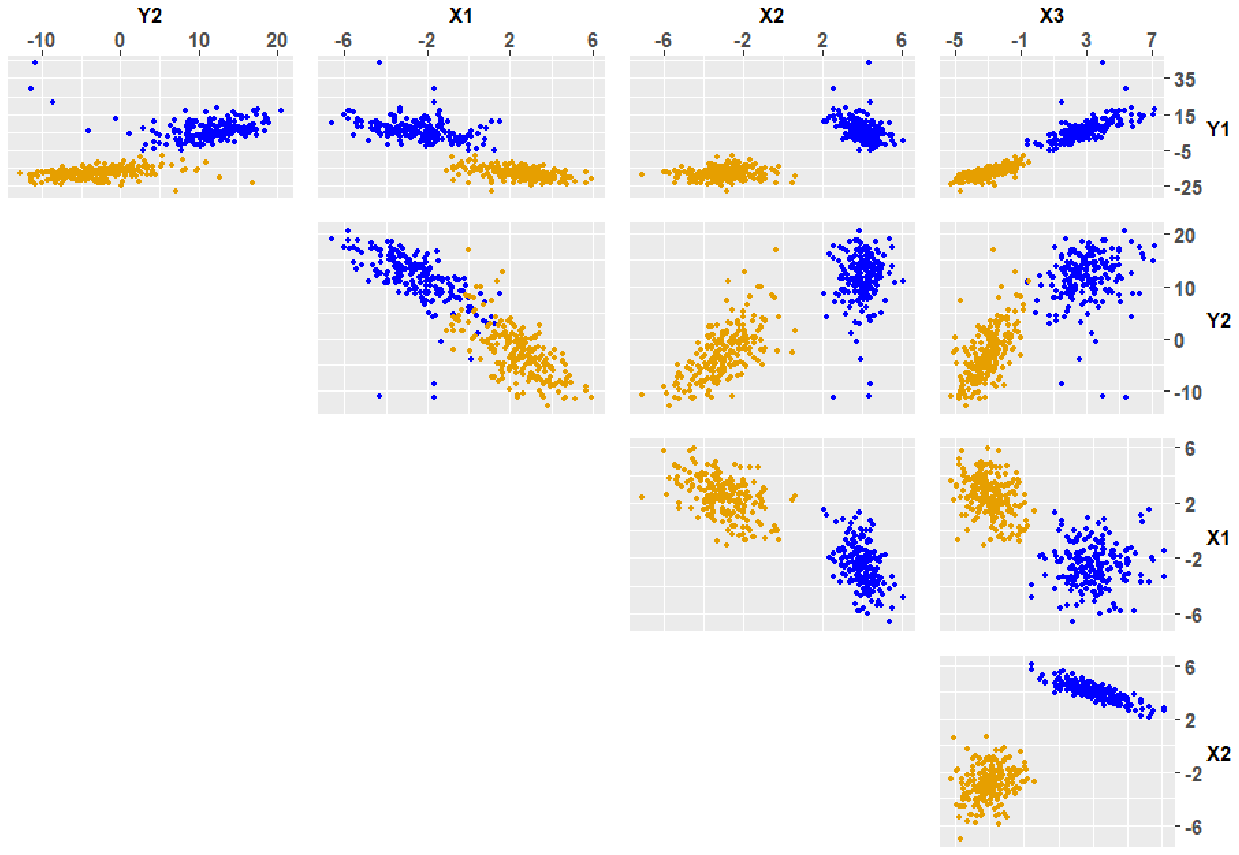}}
\caption{Pairwise plots of example datasets generated by (a) ST-ST CWM, (b) ST-N CWM and (c) N-ST CWM.
}
\label{fig:sim2}
\end{figure}

For illustrative purposes, the CWMs based on the ST distribution namely the ST-ST CWM, ST-N CWM and N-ST CWM are now considered.
These models are examples of the scenarios 1, 3 and 4 described in the previous section.
For each of these three models, 100 datasets are generated and all the CWMs, as well as the FMRs for which the distribution of the responses given the covariates is one of those considered in this manuscript. Furthermore these are fitted for $G\in\left\{1,2,3\right\}$.
We set $p = 2$, $r = 3$, $N = 400$ and the parameters displayed in~\tablename~\ref{tab:param2} to generate the datasets.
An example of generated dataset from each CWM is displayed in~\figurename~\ref{fig:sim2} and it is clear that there is a grouping structure in the covariates. 
\begin{table}[!ht]
    \centering
\caption{Parameters used to generate the simulated datasets based on the ST distribution.}
\label{tab:param2}       
       \begin{tabular}{lcc}
				\hline\noalign{\smallskip}
	Parameter  & Group 1 & Group 2 \\
	      \hline\noalign{\smallskip}
	$\pi_g$	& 0.50 & 0.50 \\		
	$\vecmu_g$	 & $(-2.50,4.00,3.00)'$  & $(2.50,-3.00,-3.00)'$    \\
	$\vecalp_{X_g}$ & $(-3.00,2.50,-2.00)'$  & $(2.50,3.00,-1.50)'$ \\
	$\matsig_{X_g}$  & $\begin{pmatrix*}[r]  2.90 & -0.50 & -0.05  \\
                                  -0.50 & 0.45 & -0.75  \\
																	-0.05 & -0.75 & 1.95  \end{pmatrix*}$ & $\begin{pmatrix*}[r]  2.30 & -0.90 & -0.35  \\
                                                                                         -0.90 & 1.55 & 0.25  \\
		 															                                                       -0.35 & 0.25 & 1.00  \end{pmatrix*}$ \\
  $\nu_{X_g}$ & 7.00 & 7.00 \\
  $\vecB_g$   & $\begin{pmatrix*}[r]  -6.00 & 1.00  \\
                                  -1.50 & -1.50 \\
																	-0.50 & 1.50 \\
																	 2.50 &	 1.50 \end{pmatrix*}$ & $\begin{pmatrix*}[r]  -10.00 & 7.50  \\
                                                                                   -1.00 & -1.00 \\
																																									  -0.50 &  1.50 \\
																																									 2.00 & 2.00 \end{pmatrix*}$\\
	$\vecalp_{Y_g}$ & $(2.00,-2.50)'$  & $(-1.00,2.00)'$ \\
	$\matsig_{Y_g}$ & $\begin{pmatrix*}[r]  1.80 & -0.30  \\
                                  -0.30 & 2.00  \end{pmatrix*}$ & $\begin{pmatrix*}[r]  2.00 & -0.35  \\
                                                                                  -0.35 & 2.80  \end{pmatrix*}$ \\
 	$\nu_{Y_g}$ &	7.00 & 7.00 \\																																						
	\noalign{\smallskip}\hline
        \end{tabular}
\end{table}

\begin{figure}[!ht]
\centering
\subfigure[\label{fig:cwm1}]
{\includegraphics[width=0.495\textwidth]{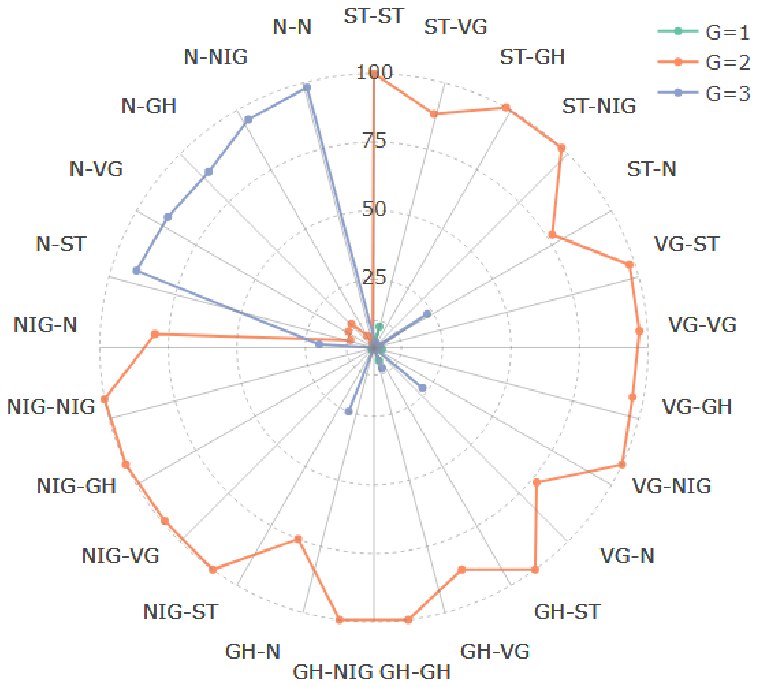}}
\subfigure[\label{fig:cwm2}]
{\includegraphics[width=0.495\textwidth]{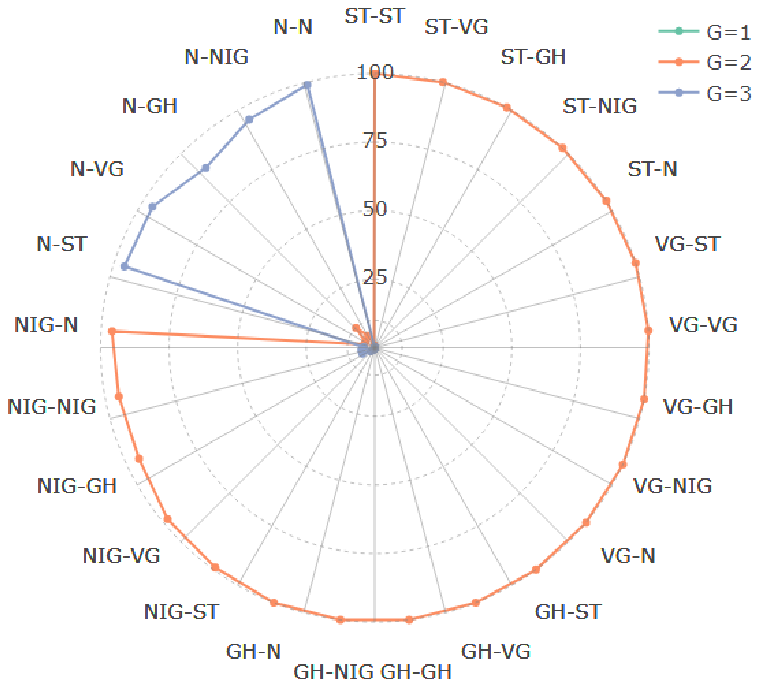}}
\subfigure[\label{fig:cwm3}]
{\includegraphics[width=0.495\textwidth]{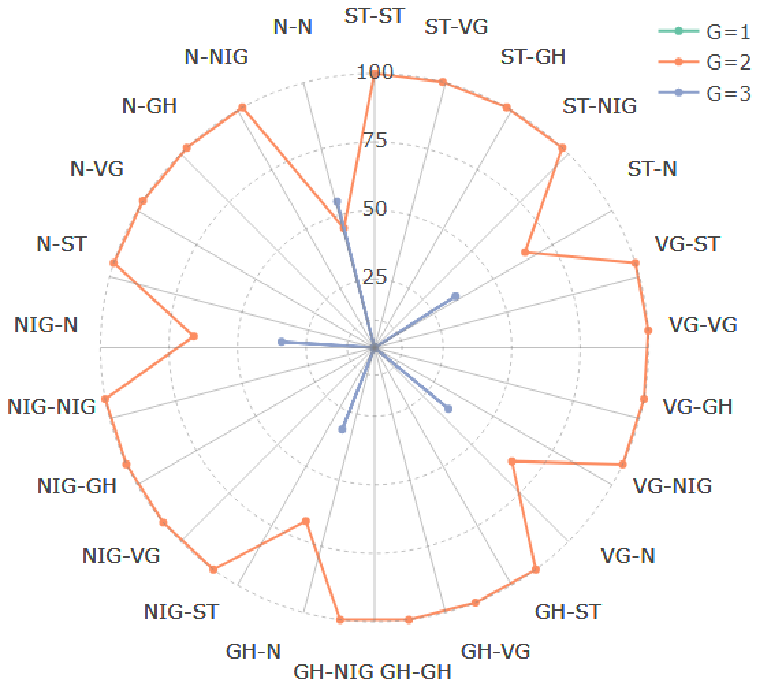}}
\caption{Radar plots of the number of times each $G$ is chosen by the BIC, for the CWMs, when the data are generated from (a) ST-ST CWM, (b) ST-N CWM and (c) N-ST CWM.
Each sub-plot refers to 100 datasets.}
\label{fig:CWM_BIC}
\end{figure}

\begin{figure}[!ht]
\centering
\subfigure[\label{fig:fmr1}]
{\includegraphics[width=0.495\textwidth]{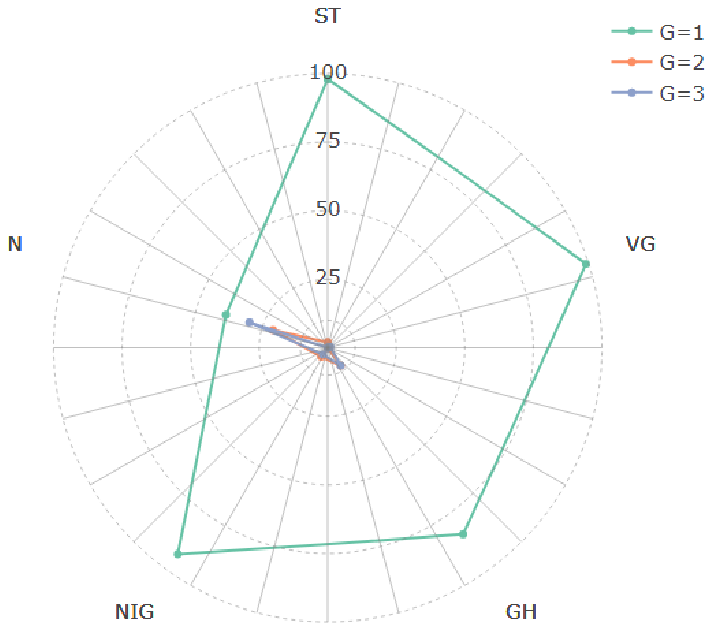}}
\subfigure[\label{fig:fmr2}]
{\includegraphics[width=0.495\textwidth]{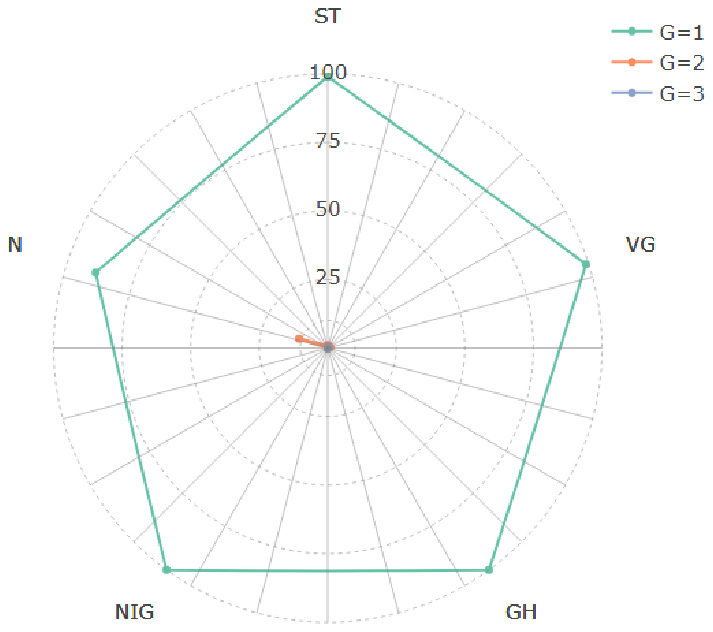}}
\subfigure[\label{fig:fmr3}]
{\includegraphics[width=0.495\textwidth]{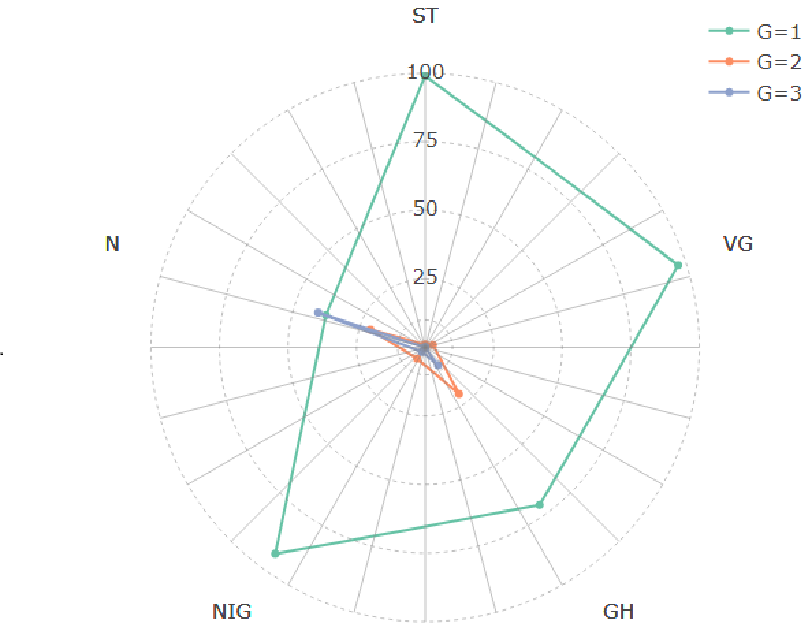}}
\caption{Radar plots of the number of times each $G$ is chosen by the BIC, for the FMRs, when the data are generated from (a) ST-ST CWM, (b) ST-N CWM and (c) N-ST CWM.
Each sub-plot refers to 100 datasets.}
\label{fig:FMR_BIC}
\end{figure}

The results are illustrated in the radar plots of~\figurename~\ref{fig:CWM_BIC} and~\figurename~\ref{fig:FMR_BIC}, for the CWMs and FMRs, respectively.
In detail, each sub-plot shows the number of times each $G$ is chosen by the BIC for each model over the 100 datasets.
Starting with the CWM results, in~\figurename~\ref{fig:cwm1} we can see that when the data are generated by the ST-ST CWM, all the CWMs for which either $p_{\vecX}$, $p_{\vecY}$, or both are assumed to be normal, problems arise in detecting the true number of groups in the data.
As discussed in Section~\ref{sec:intro}, when the normal distribution is used for modelling skewed data, it has has a tendency to over fit the true number of groups.
This is confirmed by our results, but it is also interesting to notice that this issue has a different magnitude depending on which one of $p_\vecX$ or $p_\vecY$ is modelled using the normal density.
Specifically, when $p_\vecX$ is assumed to be skewed and $p_\vecY$ assumed to be normal, most of the time $G=2$ is still properly selected, although it is still not as accurate as the CWMs where both $p_\vecX$ and $p_\vecY$ are assumed to be skewed.
On the other hand, when $p_\vecX$ is assumed normal and $p_\vecY$ assumed skewed, $G=3$ is nearly always chosen.

When the datasets are generated from an ST-N CWM, the only models having serious problems are those when the covariates are assumed to be normally distributed, as shown in~\figurename~\ref{fig:cwm2}.
Because of their greater flexibility, all the CWMs that assume a skewed density for $p_\vecY$ are able to accurately model symmetric data.
The results for the N-ST CWM are displayed in~\figurename~\ref{fig:cwm3}.
Here, the only CWMs that present issues are those for which $p_\vecY$ is assumed normal.

Regarding the capability of the BIC in detecting the exact data generating model, we observed that over the 100 datasets generated by the ST-ST and N-ST CWMs, the BIC selects the correct model 78 and 82 times, respectively.
The occasions in which it fails are due to a wrong distribution chosen for only one of the covariates or the conditional distribution of the responses.
Under no circumstances are both distributions incorrectly chosen.
When the ST-N CWM is considered, the BIC performance is even better than before, as it selects the correct model 99 times.

\begin{figure}[!ht]
\centering
\includegraphics{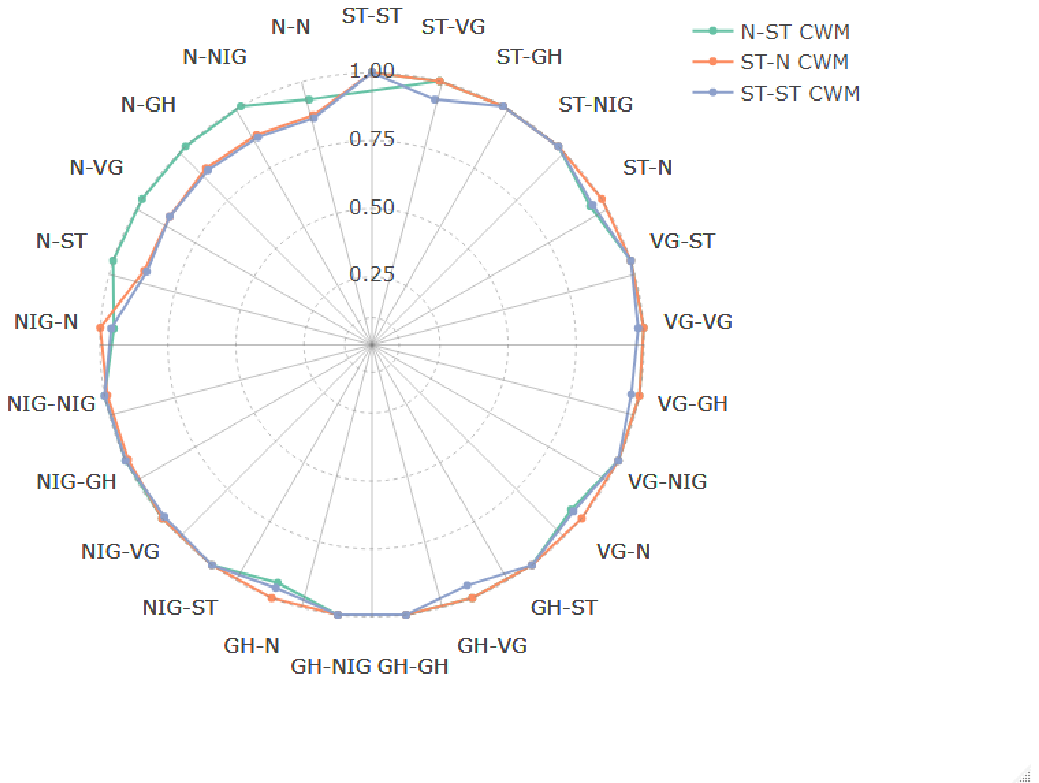}
\caption{Average ARI values of the CWMs, computed over 100 datasets for each of the three data generating models.}
\label{fig:classCWM}
\end{figure}

From the analysis of the FMR results, we can see that in all the three cases illustrated in~\Cref{fig:fmr1,fig:fmr2,fig:fmr3}, when the skewed FMRs are considered, $G=1$ is repeatedly selected. Despite the clear separation between the two groups, the FMR approach is unable to correctly identify them.
The classification results of the CWMs are shown in~\figurename~\ref{fig:classCWM}.
Here, the models that have the lowest ARI values are those assuming normal covariates for the datasets generated from the ST-N and ST-ST models.
All the other CWMs produce very good classifications for all three of the data generating models considered. 

\section{Real Data Applications}
\label{sec:real}

\subsection{Overview}
In this section, all the CWMs discussed herein, as well as the FMRs for comparison purposes, are applied to two real datasets.

\subsection{Data}
\label{sec:data}

The first application considers the \texttt{AIS} dataset included in the {\tt sn} package \citep{sn}.
It contains measurements of $N_1=102$ male and $N_2=100$ female athletes (then, $N=202$ and $G=2$) collected at the Australian Institute of Sport.
The subset of seven variables, used recently in the mixtures of regression literature (\citealp{soffritti2011multivariate, dang17}) is now analyzed. Specifically, we consider red cell count (RCC), white cell count (WCC), plasma ferritin concentration (FE), body mass index (BMI), sum of skin folds (SSF), body fat percentage (BFT), and lean body mass (LBM).
As in \citet{dang17}, the blood composition variables (RCC, WCC and FE) are selected as the response variables, while the biometrical variables (BMI, SSF, BFT and LBM) are the covariates.
For this dataset, we know the true group memberships, and can therefore evaluate the clustering results of the competing models by computing the ARI.

The second application considers the \texttt{pulpfiber} dataset included in the \textbf{robustbase} package \citep{robustbase}.
The data contains measurements related to the properties of $N=62$ pulp fibers and the resultant paper produced.
The following subset of four variables is analyzed here: elastic modulus (EM), stress at failure (SF), long fiber fraction (LFF) and zero span tensile (ZST).
The paper properties (EM and SF) are selected as the response variables, while the pulp fiber characteristics (LFF and ZST) are the covariates.
As opposed to the AIS data, the group structure is completely unknown.
Although we cannot compute the ARI to evaluate the partitions of the competing models, from the investigation of the pairwise plot, it will be quite evident the existence of a grouping structure in the data.

\subsection{Results}
\label{sec:res}

\begin{figure}[!ht]
\centering
\includegraphics[width=0.95\textwidth]{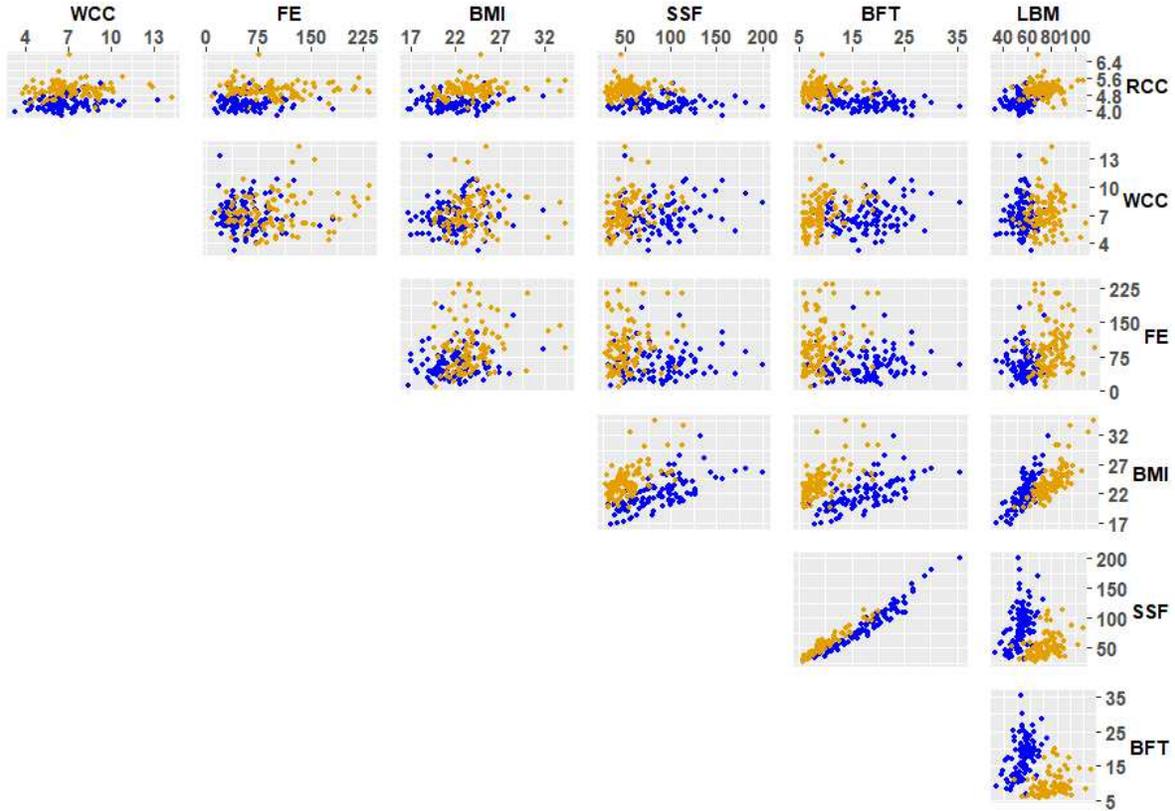}
\caption{Pairwise plot of the \texttt{AIS} dataset coloured according to the classification produced by the best fitting CWM.}
\label{fig:ais_col}
\end{figure}

In both applications, all the CWMs and the FMRs considered in this manuscript are fitted with $G\in\left\{1,2,3\right\}$.
When the \texttt{AIS} dataset is considered, the best CWM according to the BIC is the GH-ST with $G=2$, whereas the best FMR model is the VG with $G=1$.
The classification results give an ARI of 0.96 for the GH-ST CWM, i.e., an almost perfect classification, and an ARI of 0 for the VG FMR.
Our results are similar to those in \citet{dang17}, where the FMR model, based on the normal distribution, detected only one group in the data.
This means that, even if skewed distributions are used for the FMR models, they are unable to correctly model this data.
However, the GH-ST CWM performs better than the best N-N parsimonious CWM reported as 0.92 in \cite{dang17}.
This aspect can be better understood by looking at the pairwise plot of the dataset in~\figurename~\ref{fig:ais_col}, and coloured according to the classification produced by the GH-ST CWM.
As we can see, many of the variables seem to present a skewed behaviour, so that our distributions are able to model the data in a more accurate way than the normal distribution.

\begin{figure}[!ht]
\centering
\includegraphics[width=0.7\textwidth]{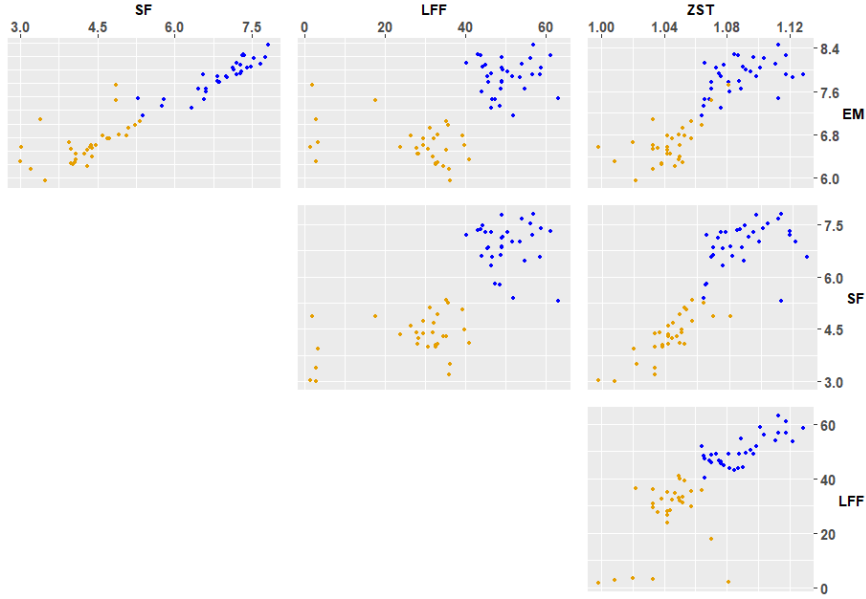}
\caption{Pairwise plot of the \texttt{pulpfiber} dataset colored according to the classification produced by the best fitting CWM.}
\label{fig:pulp_col}
\end{figure}

For the \texttt{pulpfiber} dataset, the best CWM according to the BIC is the VG-N with $G=2$, whereas the best FMR model is the ST with $G=1$.
Similar to the previous application, the best CWM detects two groups in the data, while the best FMR model only finds one group.
By looking at the pairwise plot of the dataset in~\figurename~\ref{fig:pulp_col}, coloured according to the classification produced by the VG-N CWM, it seems clear that there is more than one group in the data.
Specifically, it appears that there are two seemingly skewed and separated groups with possible mild outliers. Moreover, it is interesting to note that the linear relationship between the response variables and covariates appear to be similar between the two groups found by the CWMs, which may explain why the FMR only finds one group.

\section{Summary}
\label{sec:end}

A novel family of 24 multivariate CWMs was introduced.
Extending the completely unconstrained normal CWM of \citet{dang17}, the distributions of the responses and of the covariates were allowed to be skewed.
For illustrative purposes the following four skewed distributions were considered: the generalized hyperbolic, the skew-$t$, the variance-gamma and the normal inverse Gaussian.
Additionally, by also considering the normal distribution, our models were flexible enough to consider scenarios in which the covariates and the responses conditioned on the covariates are skewed, or in which one of the two sets of variables is normally distributed and the other one is skewed.

An EM algorithm was discussed for parameter estimation, and its capability of recovering the parameters of the data generating model was tested in a simulation study.
A comparison among the CWMs and the FMRs was also investigated via simulated data.
Specifically, it was shown that by ignoring the distribution of the covariates, the FMRs may fail to detect the correct number of groups in the data, even if they are well separated.

All our CWMs, as well as the normal CWM and the FMRs, were additionally fitted to two real datasets.
The results of the first application are in line with those present in the literature in the sense that the FMRs are not able to model this dataset, even by using skewed distributions; however, one of the skewed CWMs outperformed the classification result obtained by \cite{dang17}.

In the second application, despite lacking a true classification, an underlying group structure is evident by a graphical analysis.
In such a case, one of our CWMs seems to properly identify these groups, while the FMRs find just one group, similar to the first application.

Possible extensions of this work might be to consider a parsimonious structure for the covariance matrices, in the fashion of \citet{dang17}, as well as restraining the parameters governing the tail behaviour.

\bibliography{SkewCWM.bib}
\end{document}